# Anomalous relation between in-plane and out-of-plane stiffnesses in 2D networked materials

Fei Pan[1,†], Feng Zhang[1,†], Yuli Chen[1,2,*], Zhi Liu[3], Xiaoling Zheng[4], Bin Liu[3]

[1] Institute of Solid Mechanics, Beihang University (BUAA), Beijing 100191, China

[2] Department of Civil and Environmental Engineering, Northwestern University, Evanston, IL 60208, USA

[3] AML, CNMM, Department of Engineering Mechanics, Tsinghua University, Beijing 100084, China

[4] Shanghai Aircraft Design and Research Institute, Commercial Aircraft Corporation of China Ltd, Shanghai 201210, China

† These authors contributed equally to this work.

## Abstract:

For thin networked materials, which are spatial discrete structures constructed by continuum components, a paradox on the effective thickness defined by the in-plane and out-of-plane stiffnesses is found, *i.e.* the effective thickness is not a constant but varies with loading modes. To reveal the mechanism underneath the paradox, we have established a micromechanical framework to investigate the deformation mechanism and predict the stiffness matrix of the networked materials. It is revealed that the networked materials can carry in-plane loads by axial stretching/compression of the components in the networks and resist out-of-plane loading by bending and torsion of the components. The bending deformation of components has a corresponding relation to the axial stretching/compression through the effective thickness, as the continuum plates do, while the torsion deformation has no relation to the axial stretching/compression. The isolated torsion deformation breaks the classical stiffness relation between the in-plane stiffness and the out-of-plane stiffness, which can even be further distorted by the stiffness threshold effect in randomly networked materials. Accordingly, a new formula is summarized to describe the anomalous stiffness relation. This network model can also apply in atomic scale 2D nanomaterials when combining with the molecular structural mechanics model. This work gives an insight into the understanding of the mechanical properties of discrete materials/structures ranging from atomic scale to macro scale.

[*] **Corresponding author**: Yuli Chen, E-mail address: yulichen@buaa.edu.cn

## 1 Introduction

Thin planar networked materials constructed by regularly or randomly distributed walls/filaments, ranging from nanoscale materials like carbon nanotube films (Pan et al., 2017; Yang et al., 2013; Yu et al., 2014) and metal nanowire networks (Guo and Ren, 2015; Huang and Zhu, 2018; Ye et al., 2014), to macroscale materials, such as porous metal fiber sintered sheets (Jin et al., 2013), non-woven fabrics (Chen et al., 2016; Grandgeorge et al., 2018) and cellular thin plates (Chen et al., 2018; Lee et al., 2017), have attracted much attention due to their unique performances, such as light weight, high porosity, high in-plane rigidity and out-of-plane flexibility (Ban et al., 2016; Grandgeorge et al., 2018; Yang et al., 2018), which can be applied in many fields, such as flexible electronics (Chen et al., 2018; Son et al., 2018; Ye et al., 2018), thin plate mechanical metamaterials (Davami et al., 2015; Lin et al., 2018; Wei et al., 2018; Zhang et al., 2018), batteries electrodes (Aqil et al., 2015; Liu et al., 2016), filtering membranes (Cooper et al., 2003; Liang et al., 2018), and functional coatings (Gagné and Therriault, 2014; Kumar et al., 2015). In most applications, the networked materials endure both in-plane and out-of-plane loads.

Besides the extensively studied in-plane mechanical behavior (Ban et al., 2016; Berhan et al., 2004; Chen et al., 2015; Gibson and Ashby, 1999; Li et al., 2018; Pan et al., 2016; Picu, 2011; Wei et al., 2016), the out-of-plane mechanical properties (e.g. the bending stiffness) have also attracted considerable interest for their potential applications where the out-of-plane deformation can be harnessed (Grandgeorge et al., 2018; Huang and Zhu, 2018; Pan et al., 2017; Son et al., 2018). It is widely accepted that the out-of-plane stiffness **D** of a thin continuum plate can be easily derived from its in-plane stiffness **A** and thickness $t$, as (Timoshenko and Woinowsky-Krieger, 1959)

$$\mathbf{D} = \frac{t^2}{12}\mathbf{A}. \tag{1.1}$$

Thus there is usually no need to investigate the out-of-plane stiffness of plate-like materials, because the out-of-plane rigidity actually originates from the in-plane stretching/compression or shearing deformation for continuum plates. However, in some cases, this relation does not hold any more. For example, when going from macroscopic continuum solids into atomic scale discrete structures, e.g. 2-dimensional (2D) materials like graphene, the relation between the out-of-plane and in-plane stiffnesses would become anomalous. They no longer satisfy the relation as the continuum plates do. Even a self-consistent effective thickness cannot be found, because the thickness is dependent on loading modes (Gao and Xu, 2015; Huang et al., 2006; Peng et al., 2008; Wu et al., 2008). The main reason lies in the totally different local deformation mechanism of the discrete structures. The out-of-plane stiffness of discrete structures is from the variation of interatomic potentials, which is related to the relative positions of atoms, but not from the in-plane deformations.

As for the networked materials, they can be regarded as spatial discrete structures constructed by continuum elements, and is located in the gap between the fully discrete materials and the continuum materials, as sketched in Fig. 1. Due to the semi-discrete and semi-continuum feature, it is reasonable to guess that the out-of-plane mechanical properties of the networked materials may have some difference from the continuum plates, and thus should be carefully reexamined. Particularly, the following two questions need to be addressed:

(1) How do these materials resist an arbitrary out-of-plane deformation? Are they more like the discrete 2D materials or the continuum plates?

(2) What is the relation between in-plane stiffness and out-of-plane stiffness? Is it possible to find a self-consistent effective thickness to match the classical plate model?

To answer these questions, the out-of-plane stiffness of regular and random 2D networked materials is systematically investigated theoretically and numerically in this work.

The rest of the paper is organized as follows. First, we report a paradox on the thickness for some typical networked materials, which is similar to the 2D materials. Second, in Sections 3 and 4, a theoretical micromechanical framework is established to reveal the deformation mechanism and predict the in-plane and out-of-plane stiffnesses of the networked materials. Then, some examples on regular and random networks are presented to make a specific understanding of the in-plane and out-of-plane stiffnesses in Section 5. In Sections 6 and 7, the discussion on the mechanism of paradox and a brief conclusion are presented.

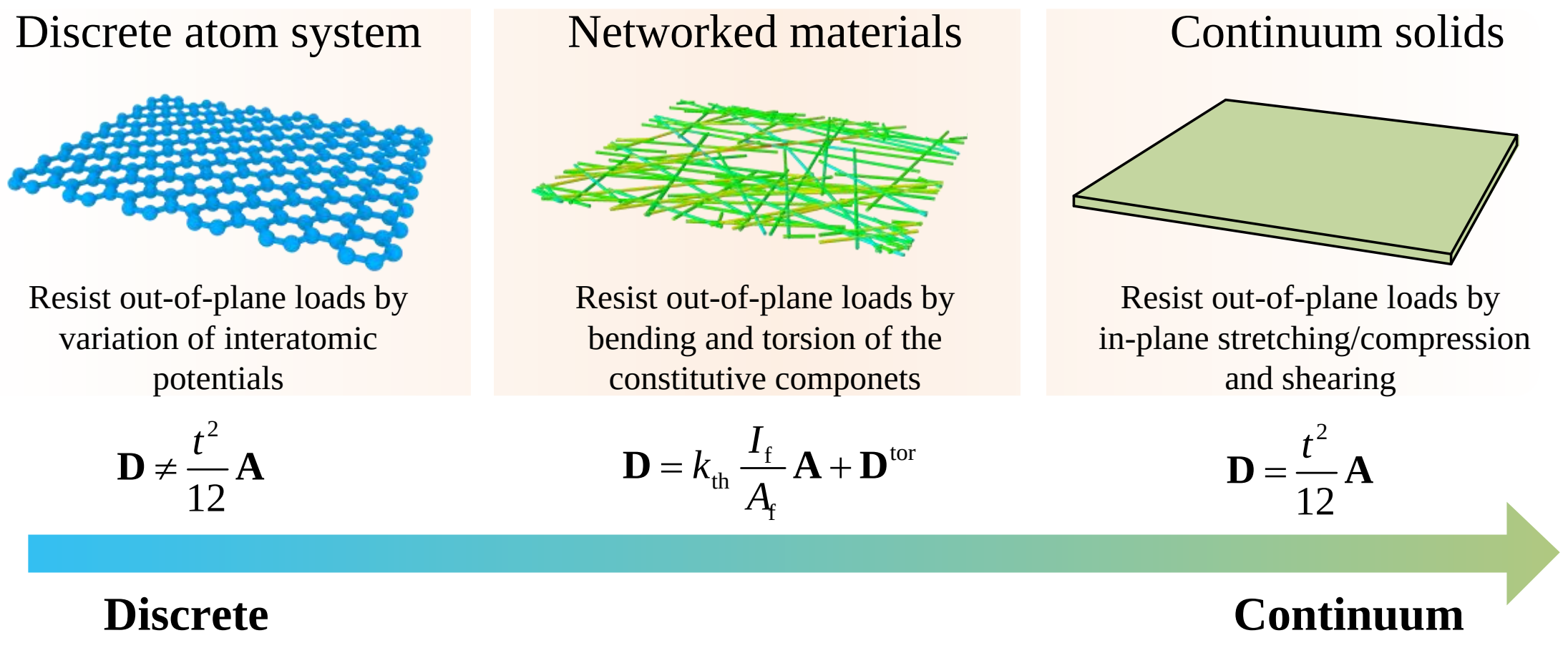

**Fig. 1** Scheme diagram of the relation between the in-plane and out-of-plane stiffnesses of 2D materials constructed by single layered discrete atom system, thin networked materials architected by contiumm components and contiumm solids.

## 2 Paradox on the thickness

For a thin plate, the elastic in-plane and out-of-plane deformations can be governed by (Reddy, 2004; Timoshenko and Woinowsky-Krieger, 1959)

$$\begin{bmatrix} \mathbf{N} \\ \mathbf{M} \end{bmatrix} = \begin{bmatrix} \mathbf{A} & \mathbf{B} \\ \mathbf{B} & \mathbf{D} \end{bmatrix} \begin{bmatrix} \boldsymbol{\varepsilon} \\ \boldsymbol{\kappa} \end{bmatrix}, \tag{2.1}$$

Here $\boldsymbol{\varepsilon} = \begin{bmatrix} \varepsilon_1 & \varepsilon_2 & \varepsilon_3 \end{bmatrix}^{\mathrm{T}}$ represents the simplified notations for the in-plane strains, in which $\varepsilon_1$ and $\varepsilon_2$ are the normal strains in directions 1 and 2, respectively, and $\varepsilon_3$ is the engineering shear strain (as shown in Fig. 3c). $\boldsymbol{\kappa} = \begin{bmatrix} \kappa_1 & \kappa_2 & \kappa_3 \end{bmatrix}^{\mathrm{T}}$ represents the out-of-plane curvatures, in which $\kappa_1$ and $\kappa_2$ are the bending curvatures along directions 1 and 2, respectively, and $\kappa_3$ is the torsional curvature (as shown in Fig. 3c). $\mathbf{N}=\left[N_1 \quad N_2 \quad N_3\right]^{\mathrm{T}}$ and $\mathbf{M}=\left[M_1 \quad M_2 \quad M_3\right]^{\mathrm{T}}$ are the corresponding sectional forces and bending/torsion moments of unit width, respectively. **A** and **D** are the in-plane and out-of-plane stiffness matrixes, respectively, and **B** is the coupling stiffness matrix.

In classical continuum plate theory, the out-of-plane stiffness **D** can be derived from the in-plane stiffness **A** and the thickness $t$ (Timoshenko and Woinowsky-Krieger, 1959), as expressed in Eq.(1.1), and thus the thickness can be obtained as

$$t=\sqrt{12\frac{D_{ij}}{A_{ij}}}\,, \tag{2.2}$$

Note here that the subscript $ij$ refers to the items of the matrixes, and Einstein's summation convention does not apply to them. The relation in Eq.(2.2) has been used as a guidance to characterize the thickness indirectly (Huang et al., 2006; Kudin et al., 2001; Wang et al., 2005; Yakobson et al., 1996).

Here we selected three typical regular lattice networked materials and three randomly networked materials with different densities (more details can be found in Sections 3-5), and calculated the thickness of the networks from finite element simulation results (see details in Appendix A) according to Eq.(2.2). To distinguish from the physical thickness of the material, we define it as mechanically effective thickness. To make a intuitively comparison with the classical continuum plate model, the mechanically effective thickness under different loading modes is normalized by the physical thickness as

$$\hat{t}_{ij}=\frac{\sqrt{12D_{ij}/A_{ij}}}{t}\,. \tag{2.3}$$

For a continuum plate, it should be 1.

Fig. 2 shows the normalized thickness of the networks. A paradox on the thickness is presented in these materials: for networks with the same type and density, the mechanical effective thickness is not a constant, but varies with respect to different loading modes (a certain $ij$ ). Thus it seems impossible to find an effective constant thickness to unify the mechanical features according to the continuum plate model.

Then the questions arise. Why the mechanical effective thickness cannot be described by a constant thickness? How to find the relation between the in-plane and out-of-plane stiffnesses so that we can design the structure and density of networks for a given loading mode?

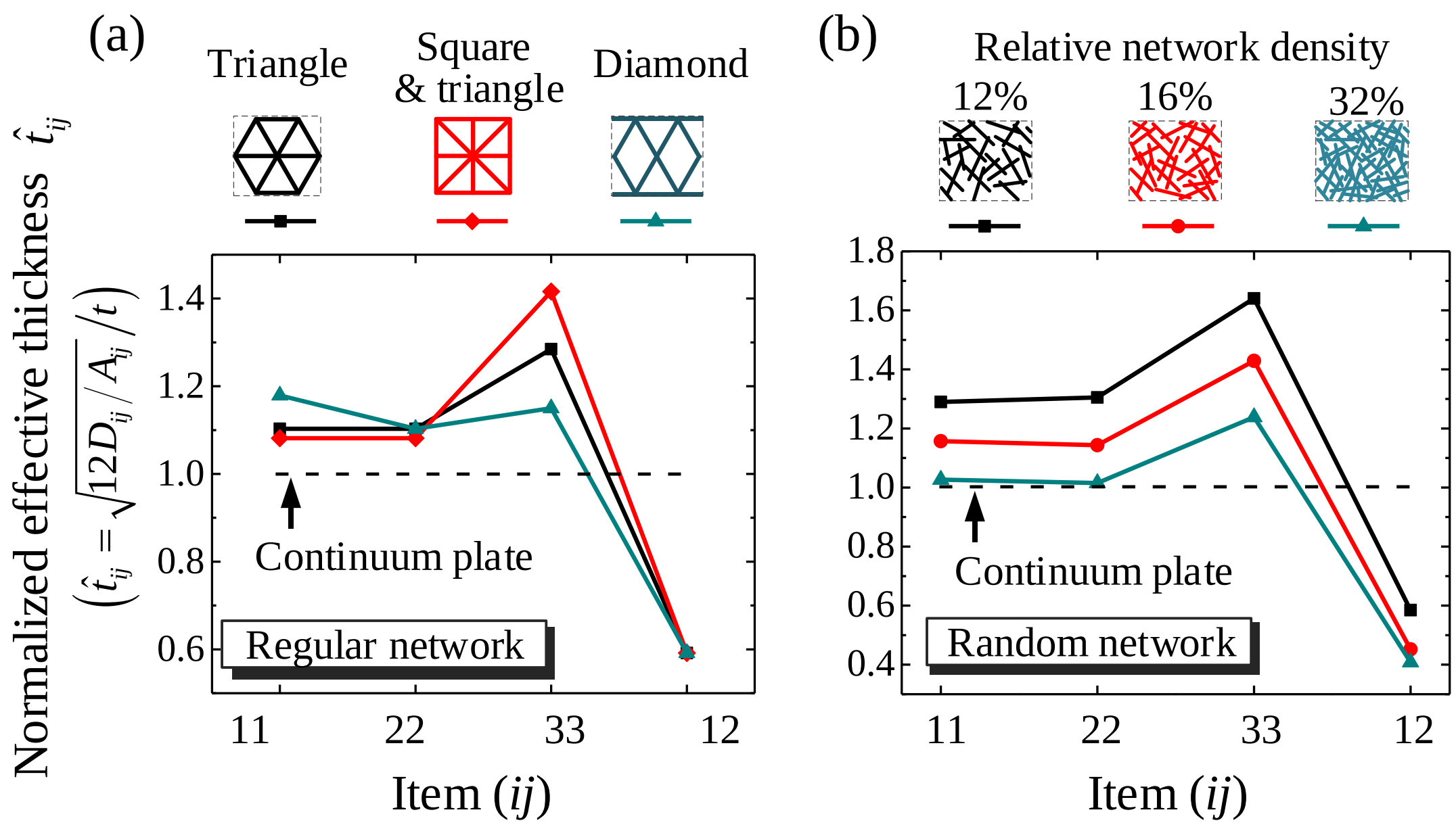

**Fig. 2** Normalized mechanically effective thickness under different loading modes calculated from finite element simulation for (a) three typical regular lattice networks and (b) three randomly distributed networks with different relative network density.

To reveal the mechanism underneath the paradox, we will establish a micromechanical model to study the in-plane and out-of-plane deformation of the networked materials, and predict the stiffness systematically.

## 3 Network model

The networked materials are architected by regularly or randomly distributed slender components (filaments or walls), as sketched in Fig. 3. Due to the extremely thin thickness, these components are assumed to lie in the same plane and penetrate into each other at the intersections, that is, the physical thickness of the network is identical to the thickness of the components, as shown in Fig. 3a and the components are rigidly connected to each other at the intersections.

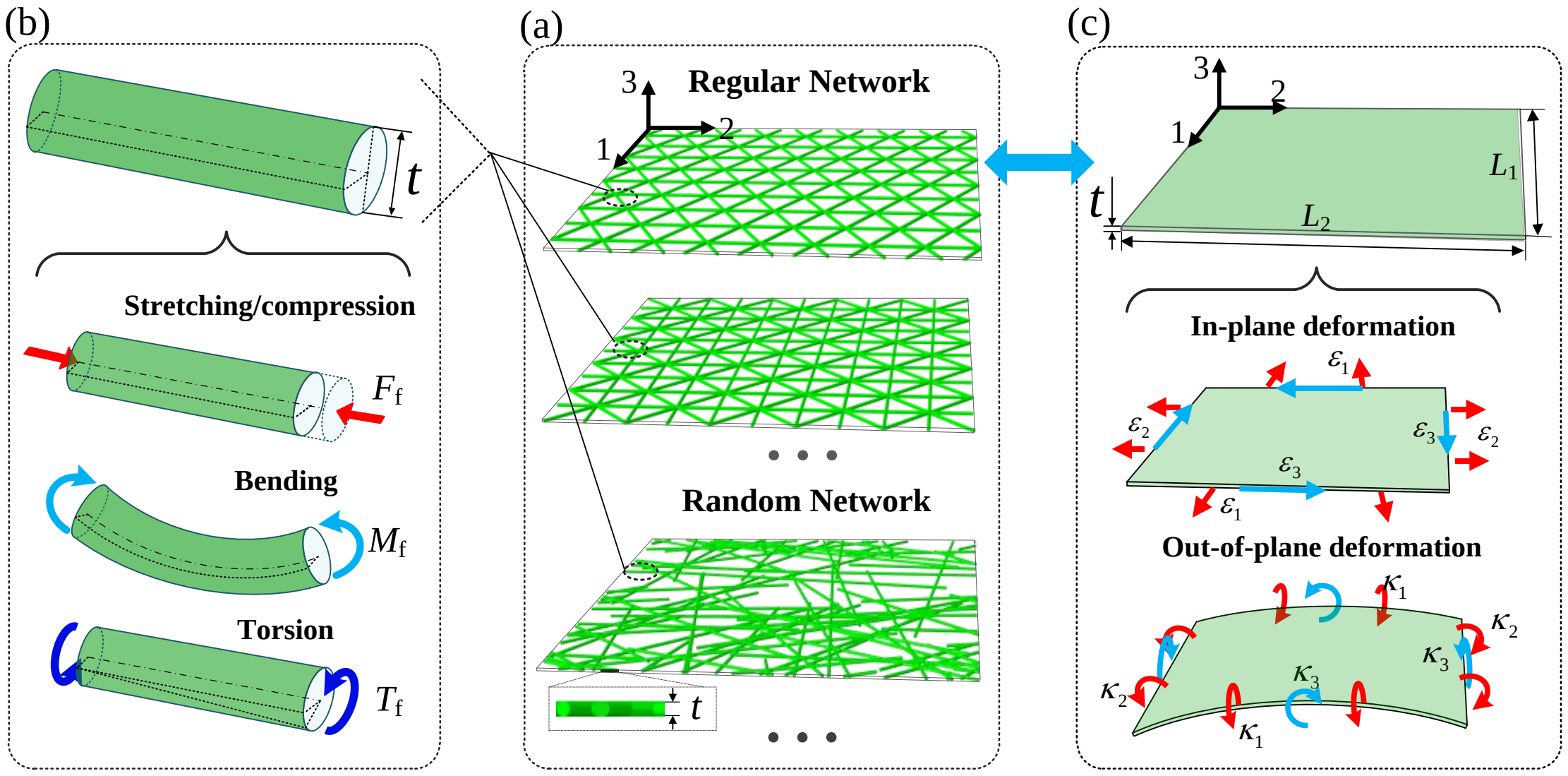

**Fig. 3** Scheme diagram of the networked materials. (a) Two typical regular networks and one random network; (b) The constitutive component of network can be simplified as beam model with three basic deformation modes, *i.e.* axial stretching/compression, transverse bending and torsion; (c) The network can be equivalent to a thin plate model with in-plane and out-of-plane deformations.

The relative density of the network is defined by the projection area fraction of the constitutive components, as

$$\hat{\rho} = \frac{1}{L_1 L_2} \sum l_i b_i \, , \tag{3.1}$$

where $l_i$ and $b_i$ are the length and width of the $i^{\text{th}}$ components, and $L_1$ and $L_2$ are the edge lengths of the representative area element (RAE).

As shown in Fig. 3b, the mechanical behavior of the components can be described as a beam model, whose tension, bending and torsion satisfy

$$F_{\mathrm{f}} = (EA)_{\mathrm{f}}\, \varepsilon_{\mathrm{f}} , \tag{3.2}$$

$$M_{\mathrm{f}} = (EI)_{\mathrm{f}}\, \kappa_{\mathrm{f}} , \tag{3.3}$$

$$T_{\mathrm{f}} = (GJ)_{\mathrm{f}}\, \tau_{\mathrm{f}} , \tag{3.4}$$

where $(EA)_{\mathrm{f}}$, $(EI)_{\mathrm{f}}$ and $(GJ)_{\mathrm{f}}$ are the tensile, bending and torsion stiffness, respectively; $F_{\mathrm{f}}$, $M_{\mathrm{f}}$ and $T_{\mathrm{f}}$ are tension force, bending moment and torsion moment, respectively; $\varepsilon_{\mathrm{f}}$, $\kappa_{\mathrm{f}}$ and $\tau_{\mathrm{f}}$ are axial strain, bending curvature and torsion curvature, respectively. To distinguish the in-plane and out-of-plane bending, the superscripts “in” and “out” for the items in Eq.(3.3) will be used, respectively.

Because the dimension in the direction of thickness is far less than that in the other two directions, i.e., $t \ll L_1, L_2$, the planar network can be equivalent to a thin plate, as shown in Fig. 3c, and the mechanical behavior of the equivalent thin plate can be described by Eq.(2.1).

## 4 Stiffness matrix of the network

To reveal the deformation mechanism of the networked materials, and probe the relation between the in-plane and out-of-plane stiffnesses, we will systematically study the in-plane and out-of-plane mechanical behaviors of regular and random networks, and establish a unified theoretical framework for their elasticity.

### 4.1 Deformation of the components

When under in-plane or out-of-plane load, the components have to deform collaboratively to carry the external load. Thus, we will first understand the deformation mechanism of the components.

1) In-plane deformation

It has been found that when the networked materials are under in-plane loading, the external load is mainly carried by the axial stretching/compression or in-plane bending of the constituent components (Gibson and Ashby, 1999). When the deformation of the components is dominated by stretching, *i.e.* the deformation is affine (Chen et al., 2015; Head et al., 2003), the axial stretching strain $\varepsilon_{\mathrm{f}}$ can be obtained by the Mohr's circle (Cox, 1952), as

$$\varepsilon_{\mathrm{f}} = \varepsilon_1 \cos^2\theta + \varepsilon_2 \sin^2\theta + \varepsilon_3 \sin\theta\cos\theta, \tag{4.1}$$

where $\theta$ is the angle between the component axis and direction 1. However, when the components have bending-dominated deformation or both axial stretching and transverse bending deformation, *i.e.* the deformation is non-affine, the axial strain $\varepsilon_{\mathrm{f}}$ does not have the relation in Eq.(4.1). Thus the deformation of the components (*i.e.* $\varepsilon_{\mathrm{f}}$ and $\kappa_{\mathrm{f}}^{\mathrm{in}}$) should be analyzed according to the specific constructive structure (Gibson and Ashby, 1999), or be further modified based on the affine assumption (Chen et al., 2015; Pan et al., 2016).

2) Out-of-plane deformation

An arbitrary small out-of-plane deformation of a network can be decomposed into two basic modes, two uniaxial pure bending deformations $\kappa_1$ and $\kappa_2$, and a pure torsion deformation $\kappa_3$. It has been found that, under a uniaxial pure bending deformation $\kappa_1$ (or $\kappa_2$), the components of the network can be bent and twisted simultaneously to resist the uniaxial pure bending loading on the network (Pan et al., 2017), as illustrated in Fig. 4a, and the out-of-plane bending and torsion deformations of the components, $\kappa_{\mathrm{f}}^{\mathrm{out}}$ and $\tau_{\mathrm{f}}$, can be obtained by mapping a line to a spiral on a cylindrical surface, as

$$\kappa_{\mathrm{f}}^{\mathrm{out}} = \kappa_1 \cos^2\theta, \quad \text{or} \quad \kappa_{\mathrm{f}}^{\mathrm{out}} = \kappa_2 \sin^2\theta, \tag{4.2}$$

$$\tau_{\mathrm{f}} = -\kappa_1 \sin\theta\cos\theta, \quad \text{or} \quad \tau_{\mathrm{f}} = \kappa_2 \sin\theta\cos\theta \tag{4.3}$$

Similarly, to obtain the deformation of component in a network under a pure torsion deformation $\kappa_3$, we can map the straight line to a spatial curve, as illustrated in Fig. 4b. After a small torsion, the network deforms from a plane surface to a saddle surface, which is expressed as

$$x_3 = -\frac{\kappa_3}{2} x_1 x_2 . \tag{4.4}$$

Selecting the line $\left(s\cos\theta + x_1^0, s\sin\theta + x_2^0, 0\right)$ to describe a component with an angle $\theta$ with respect to direction 1, after a small torsion deformation it becomes

$$\begin{cases} x_1 = s\cos\theta + x_1^0 \\ x_2 = s\sin\theta + x_2^0 \\ x_3 = \left(-\kappa_3/2\right)\left(s\cos\theta + x_1^0\right)\left(s\sin\theta + x_2^0\right) \end{cases}, \tag{4.5}$$

where the parameter $s$ denotes the arc length. The torsion curvature of the component can be described by the geodesic torsion in differential geometry of the curve on the saddle surface (Kreyszig, 1991) (more details can be found in Appendix B), as

$$\kappa_{\mathrm{f}}^{\mathrm{out}} = \kappa_3 \sin\theta\cos\theta, \tag{4.6}$$

$$\tau_{\mathrm{f}} = \frac{\kappa_3}{2}\cos 2\theta . \tag{4.7}$$

Therefore, by superposition, the deformation of the components in a network under arbitrary out-of-plane deformation can be obtained from Eqs. (4.2), (4.3), (4.6) and (4.7), as

$$\kappa_{\mathrm{f}}^{\mathrm{out}} = \kappa_1 \cos^2\theta + \kappa_2 \sin^2\theta + \kappa_3 \sin\theta\cos\theta, \tag{4.8}$$

$$\tau_{\mathrm{f}} = -\frac{\left(\kappa_1 - \kappa_2\right)}{2}\sin 2\theta + \frac{\kappa_3}{2}\cos 2\theta . \tag{4.9}$$

which is identical to the Mohr's circle of curvature (Guest and Pellegrino, 2006).

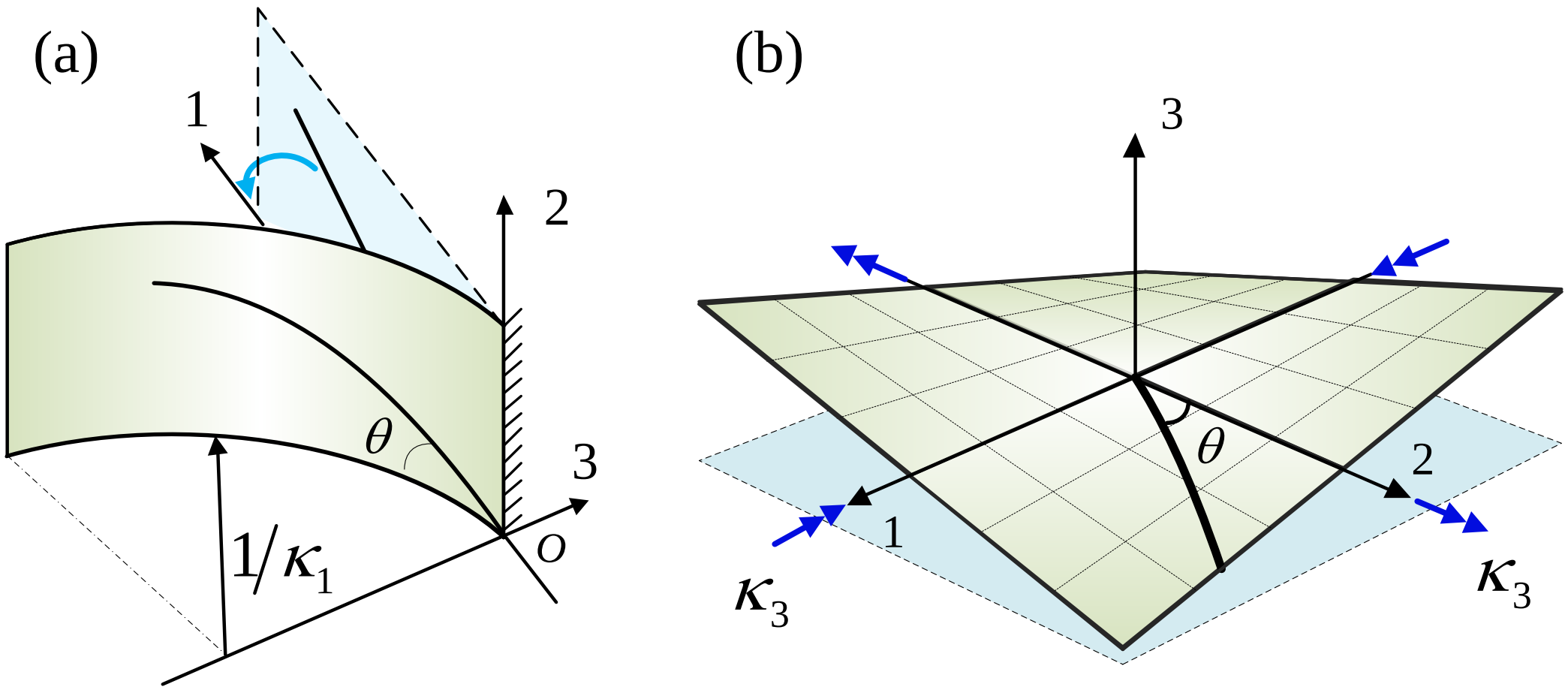

**Fig. 4** Scheme diagram for the out-of-plane deformation of a slender component in the network when under (a) a pure bending and (b) a pure torsion.

As illustrated in Section 3, the slender components in the extremely thin networked materials mainly have three basic deformation modes, *i.e.* axial stretching/compression, bending and torsion. Here we ignore the coupling between in-plane and out-of-plane deformation, *i.e.* **B**=0, the in-plane stiffness matrix **A** is induced by the axial and in-plane bending deformations while the out-of-plane stiffness matrix **D** originates from the torsion and out-of-plane bending deformations of the components. In the following, we will investigate the contributions of these deformation modes to the stiffness matrix.

## 4.2 In-plane and out-of-plane stiffnesses

As two typical in-plane deformation modes, the axial stretching/compression and the in-plane bending can both exist in the network and usually cannot be decoupled easily. Axial stretching/compression, by contrast, is more efficient on load-carrying and thus is preferred in structural materials. To make the network stretching-dominated, increasing the edge connectivity (*i.e.* the component numbers connected to one intersection) of regular network (Deshpande et al., 2001) and increasing the network density of random network (Chen et al., 2015) are found to be effective ways. Therefore, in this work we mainly focus on the stretching-dominated networks, which have relative large edge connectivity (for regular network) or network density (for random network).

In the stretching-dominated networks, we only take into account the axial stretching/compression, out-of-plane bending and torsion of the components. Therefore, the strain energy of component per unit length is expressed as

$$\bar{U}(\theta)=\frac{1}{2}\left(\left(EA\right)_{\mathrm{f}}\left(\varepsilon_{\mathrm{f}}\right)^{2}+\left(EI\right)_{\mathrm{f}}\left(\kappa_{\mathrm{f}}^{\mathrm{out}}\right)^{2}+\left(GJ\right)_{\mathrm{f}}\left(\tau_{\mathrm{f}}\right)^{2}\right). \tag{4.10}$$

For regular networks with discrete distributed components, the total strain energy of the RVE with the area $L_1 \times L_2$ can be obtained as

$$U=\sum_{i=1}^{N}\bar{U}(\theta_i)l_i\,, \tag{4.11}$$

where $N$ is the number of the components in the RVE. $l_i$ and $\theta_i$ are the length and orientation of $i^{\mathrm{th}}$ component.

For random networks, assume that the angle $\theta$ of unit length of components follows a continuous distribution function $f(\theta)$ in the range of [-$\pi$, $\pi$), and the total strain energy of the RVE can be integrated as

$$U=l_{\mathrm{total}}\int_{-\pi}^{\pi}\bar{U}(\theta)\,f(\theta)\mathrm{d}\theta\,, \tag{4.12}$$

where $l_{\mathrm{total}}=\sum l_i$ is the total length of components in the RVE.

The stiffness coefficients $A_{ij}$ and $D_{ij}$ can then be derived as

$$A_{ij}=\frac{1}{L_1L_2}\frac{\partial^2 U}{\partial\varepsilon_i\partial\varepsilon_j}\,, \tag{4.13}$$

$$D_{ij}=\frac{1}{L_1L_2}\frac{\partial^2 U}{\partial\kappa_i\partial\kappa_j}\,. \tag{4.14}$$

According to Eqs. (4.1) and (4.8)-(4.14), the stiffness coefficients $A_{ij}$ and $D_{ij}$ can be obtained analytically.

(1) For the regular distributed networks, the in-plane stiffness coefficients $A_{ij}$ are expressed as

$$A_{11} = \frac{(EA)_{\mathrm{f}}}{L_1 L_2} \sum_{i=1}^{N} l_i \cos^4 \theta_i \, , \tag{4.15}$$

$$A_{22} = \frac{(EA)_{\mathrm{f}}}{L_1 L_2} \sum_{i=1}^{N} l_i \sin^4 \theta_i \, , \tag{4.16}$$

$$A_{33} = \frac{(EA)_{\mathrm{f}}}{L_1 L_2} \sum_{i=1}^{N} \left( l_i \sin^2 \theta_i \cos^2 \theta_i \right), \tag{4.17}$$

$$A_{23} = A_{32} = \frac{(EA)_{\mathrm{f}}}{L_1 L_2} \sum_{i=1}^{N} \left( l_i \sin^3 \theta_i \cos \theta_i \right), \tag{4.18}$$

$$A_{31} = A_{13} = \frac{(EA)_{\mathrm{f}}}{L_1 L_2} \sum_{i=1}^{N} \left( l_i \sin \theta_i \cos^3 \theta_i \right), \tag{4.19}$$

$$A_{12} = A_{21} = \frac{(EA)_{\mathrm{f}}}{L_1 L_2} \sum_{i=1}^{N} \left( l_i \sin^2 \theta_i \cos^2 \theta_i \right), \tag{4.20}$$

and the out-of-plane stiffness coefficients $D_{ij}$ are summarized as

$$D_{11} = \frac{1}{L_1 L_2} \sum_{i=1}^{N} l_i \left( (EI)_{\mathrm{f}} \cos^4 \theta_i + (GJ)_{\mathrm{f}} \sin^2 \theta_i \cos^2 \theta_i \right), \tag{4.21}$$

$$D_{22} = \frac{1}{L_1 L_2} \sum_{i=1}^{N} l_i \left( (EI)_{\mathrm{f}} \sin^4 \theta_i + (GJ)_{\mathrm{f}} \sin^2 \theta_i \cos^2 \theta_i \right), \tag{4.22}$$

$$D_{33} = \frac{1}{L_1 L_2} \sum_{i=1}^{N} l_i \left( (EI)_{\mathrm{f}} \sin^2 \theta_i \cos^2 \theta_i + (GJ)_{\mathrm{f}} \frac{\cos^2 2\theta_i}{4} \right), \tag{4.23}$$

$$D_{23} = D_{32} = \frac{1}{L_1 L_2} \sum_{i=1}^{N} l_i \left( (EI)_{\mathrm{f}} \sin^3 \theta_i \cos \theta_i + (GJ)_{\mathrm{f}} \frac{\sin 2\theta_i \cos 2\theta_i}{4} \right), \tag{4.24}$$

$$D_{31} = D_{13} = \frac{1}{L_1 L_2} \sum_{i=1}^{N} l_i \left( (EI)_{\mathrm{f}} \sin \theta_i \cos^3 \theta_i - (GJ)_{\mathrm{f}} \frac{\sin 2\theta_i \cos 2\theta_i}{4} \right), \tag{4.25}$$

$$D_{12} = D_{21} = \frac{1}{L_1 L_2} \sum_{i=1}^{N} l_i \left( (EI)_{\mathrm{f}} \sin^2 \theta_i \cos^2 \theta_i - (GJ)_{\mathrm{f}} \sin^2 \theta_i \cos^2 \theta_i \right). \tag{4.26}$$

(2) For the randomly distributed networks, the in-plane stiffness coefficients $A_{ij}$ should be integrated as

$$A_{11} = \frac{l_{\text{total}}}{L_1 L_2} (EA)_{\text{f}} \int_{-\pi}^{\pi} \cos^4 \theta f(\theta) \text{d}\theta , \tag{4.27}$$

$$A_{22} = \frac{l_{\text{total}}}{L_1 L_2} (EA)_{\text{f}} \int_{-\pi}^{\pi} \sin^4 \theta f(\theta) \text{d}\theta , \tag{4.28}$$

$$A_{33} = \frac{l_{\text{total}}}{L_1 L_2} (EA)_{\text{f}} \int_{-\pi}^{\pi} \sin^2 \theta \cos^2 \theta f(\theta) \text{d}\theta , \tag{4.29}$$

$$A_{23} = A_{32} = \frac{l_{\text{total}}}{L_1 L_2} (EA)_{\text{f}} \int_{-\pi}^{\pi} \sin^3 \theta \cos \theta f(\theta) \text{d}\theta , \tag{4.30}$$

$$A_{31} = A_{13} = \frac{l_{\text{total}}}{L_1 L_2} (EA)_{\text{f}} \int_{-\pi}^{\pi} \sin \theta \cos^3 \theta f(\theta) \text{d}\theta , \tag{4.31}$$

$$A_{12} = A_{21} = \frac{l_{\text{total}}}{L_1 L_2} (EA)_{\text{f}} \int_{-\pi}^{\pi} \sin^2 \theta \cos^2 \theta f(\theta) \text{d}\theta , \tag{4.32}$$

and the out-of-plane stiffness coefficients $D_{ij}$ are integrated as

$$D_{11} = \frac{l_{\text{total}}}{L_1 L_2} \int_{-\pi}^{\pi} \left( (EI)_{\text{f}} \cos^4 \theta + (GJ)_{\text{f}} \sin^2 \theta \cos^2 \theta \right) f(\theta) \text{d}\theta , \tag{4.33}$$

$$D_{22} = \frac{l_{\text{total}}}{L_1 L_2} \int_{-\pi}^{\pi} \left( (EI)_{\text{f}} \sin^4 \theta + (GJ)_{\text{f}} \sin^2 \theta \cos^2 \theta \right) f(\theta) \text{d}\theta , \tag{4.34}$$

$$D_{33} = \frac{l_{\text{total}}}{L_1 L_2} \int_{-\pi}^{\pi} \left( (EI)_{\text{f}} \sin^2 \theta \cos^2 \theta + (GJ)_{\text{f}} \frac{\cos^2 2\theta}{4} \right) f(\theta) \text{d}\theta , \tag{4.35}$$

$$D_{23} = D_{32} = \frac{l_{\text{total}}}{L_1 L_2} \int_{-\pi}^{\pi} \left( (EI)_{\text{f}} \sin^3 \theta \cos \theta + (GJ)_{\text{f}} \frac{\sin 2\theta \cos 2\theta}{4} \right) f(\theta) \text{d}\theta , \tag{4.36}$$

$$D_{31} = D_{13} = \frac{l_{\text{total}}}{L_1 L_2} \int_{-\pi}^{\pi} \left( (EI)_{\text{f}} \sin \theta \cos^3 \theta - (GJ)_{\text{f}} \frac{\sin 2\theta \cos 2\theta}{4} \right) f(\theta) \text{d}\theta , \tag{4.37}$$

$$D_{12} = D_{21} = \frac{l_{\text{total}}}{L_1 L_2} \int_{-\pi}^{\pi} \left( (EI)_{\text{f}} \sin^2 \theta \cos^2 \theta - (GJ)_{\text{f}} \sin^2 \theta \cos^2 \theta \right) f(\theta) \text{d}\theta . \tag{4.38}$$

## 5 In-plane and out-of-plane stiffnesses of some typical networks

According to the framework in Sections 4, we can obtain the in-plane and out-of-plane stiffnesses for a given network structure. In this section, some typical examples on regular and random networks are presented to make a specific understanding of the in-plane and out-of-plane stiffnesses.

### 5.1 Regular networks

According to the former studies on the stretching-dominated regular lattices and the topological criterion proposed by Deshpande et al. (2001), when the edge connectivity is larger than or equal to 6, the network is stretching-dominated. Three typical regular lattice networked materials, triangle, mixed square and triangle, and diamond lattice networks (Wang and McDowell, 2004), are adopted here for illustration, as shown in Fig. 5. Here the length of the horizontal components is $l$. The sectional profile of the components is taken as a rectangle with the thickness $t$ and the width $b$, as shown in Fig. 5d.

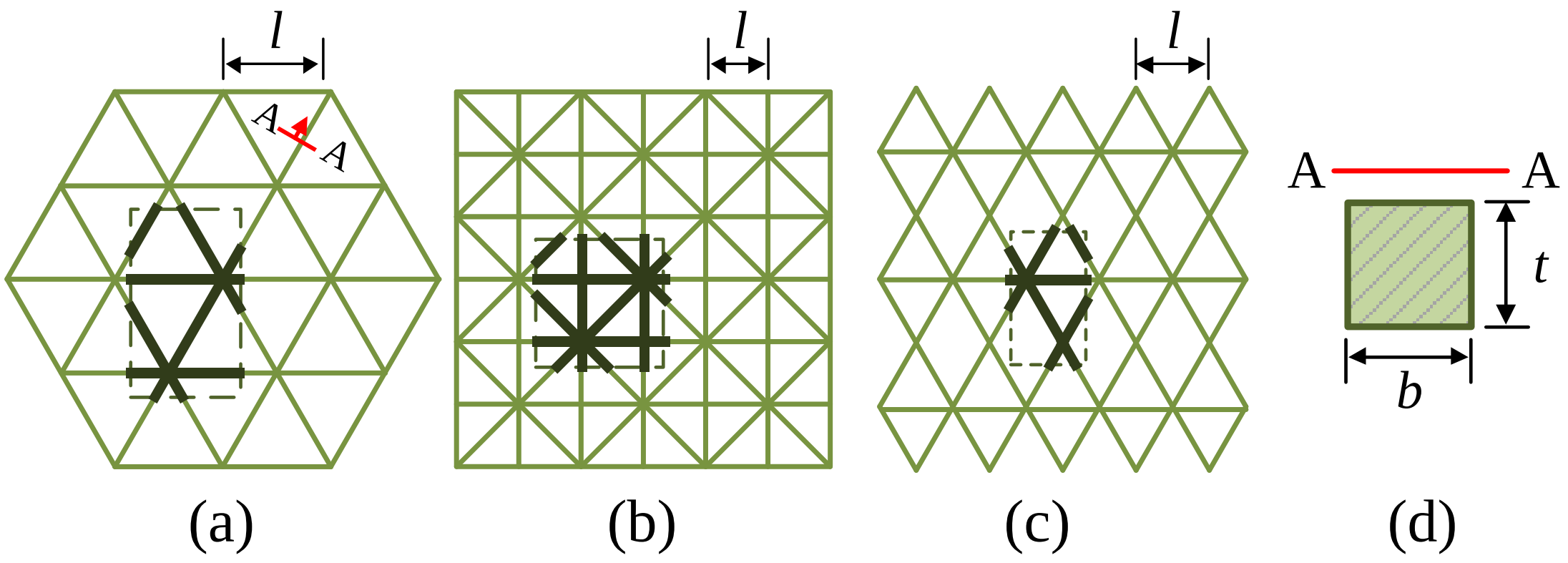

**Fig. 5** Scheme diagram of (a) triangle, (b) mixed square and triangle and (c) diamond lattice networks, as well as (d) the sectional profile of components for all the networks.

As shown in Fig. 5, the periodical RAEs are chosen in the networks. The network densities of the three regular networks can be obtained from Eq. (3.1), as listed in Table 1. Based on these RAEs, the stiffness matrixes of the three regular networks can be obtained from Eqs. (4.15)-(4.26) and Table 1, and expressed as follows.

(a) Triangle lattice network

$$\mathbf{A}=\frac{\sqrt{3}}{4l}(EA)_{\mathrm{f}}\begin{bmatrix}3 & 1 & 0\\ & 3 & 0\\ \mathrm{sym} & & 1\end{bmatrix}, \tag{5.1}$$

$$\mathbf{D}=\frac{\sqrt{3}}{4l}\begin{bmatrix}3(EI)_{\mathrm{f}}+(GJ)_{\mathrm{f}} & (EI)_{\mathrm{f}}-(GJ)_{\mathrm{f}} & 0\\ & 3(EI)_{\mathrm{f}}+(GJ)_{\mathrm{f}} & 0\\ \mathrm{sym} & & (EI)_{\mathrm{f}}+(GJ)_{\mathrm{f}}\end{bmatrix}. \tag{5.2}$$

(b) Mixed square and triangle lattice network

$$\mathbf{A}=\frac{\sqrt{2}}{4l}(EA)_{\mathrm{f}}\begin{bmatrix}2\sqrt{2}+1 & 1 & 0\\ & 2\sqrt{2}+1 & 0\\ \mathrm{sym} & & 1\end{bmatrix}, \tag{5.3}$$

$$\mathbf{D}=\frac{\sqrt{2}}{4l}\cdot\begin{bmatrix}\left(2\sqrt{2}+1\right)(EI)_{\mathrm{f}}+(GJ)_{\mathrm{f}} & (EI)_{\mathrm{f}}-(GJ)_{\mathrm{f}} & 0\\ & \left(2\sqrt{2}+1\right)(EI)_{\mathrm{f}}+(GJ)_{\mathrm{f}} & 0\\ \mathrm{sym} & & (EI)_{\mathrm{f}}+\sqrt{2}(GJ)_{\mathrm{f}}\end{bmatrix}. \tag{5.4}$$

(c) Diamond lattice network

$$\mathbf{A}=\frac{\sqrt{3}}{4l}(EA)_{\mathrm{f}}\begin{bmatrix}5/3 & 1 & 0\\ & 3 & 0\\ \mathrm{sym} & & 1\end{bmatrix}, \tag{5.5}$$

$$\mathbf{D}=\frac{\sqrt{3}}{4l}\begin{bmatrix}5/3(EI)_{\mathrm{f}}+(GJ)_{\mathrm{f}} & (EI)_{\mathrm{f}}-(GJ)_{\mathrm{f}} & 0\\ & 3(EI)_{\mathrm{f}}+(GJ)_{\mathrm{f}} & 0\\ \mathrm{sym} & & (EI)_{\mathrm{f}}+2/3(GJ)_{\mathrm{f}}\end{bmatrix}. \tag{5.6}$$

Note that the coupling stiffness matrix **B**=0 holds for all of the three networks. It can be found that the torsion of the component has a significant influence on the out-of-plane stiffness, and especially, it decreases the coupling coefficient $D_{12}$.

To verify this theoretical model, literature results and finite element method (FEM) simulation are employed here.

### 5.1.1 *In-plane stiffness*

According to Eqs. (5.1), (5.3) and (5.5), it can be found that the in-plane mechanical properties of the three lattice networks can be regarded as isotropic, special orthotropic with $E_1 = E_2$ and orthotropic materials, respectively. Based on the stiffness matrix, the effective elastic constants can be derived as (Timoshenko and Woinowsky-Krieger, 1959)

$$\nu_{12} = \frac{A_{12}}{A_{11}} \text{ and } \nu_{21} = \frac{A_{21}}{A_{22}}, \tag{5.7}$$

$$E_1 = \frac{A_{11}}{t}\left(1-\nu_{12}\nu_{21}\right), \tag{5.8}$$

$$E_2 = \frac{A_{22}}{t}\left(1-\nu_{12}\nu_{21}\right), \tag{5.9}$$

$$G_{12} = \frac{A_{33}}{t}. \tag{5.10}$$

where $\nu_{12}$ and $\nu_{21}$ are Poisson's ratio, $E_1$, $E_2$ are elastic modulus and $G_{12}$ is shear modulus. The elastic constants of the three networks are listed in Table 1, where the material of the components is taken as isotropic with the elastic modulus $E_f$. The results match well with the results obtained by Wang and Mcdowell's method (Wang and McDowell, 2004).

**Table 1** Relative density and in-plane elastic constants of three typical lattice networks

| Network | $\hat{\rho}$ | $E_1/E_f$ | $E_2/E_f$ | $G_{12}/E_f$ | $\nu_{12}$ | $\nu_{21}$ |
|---|---|---|---|---|---|---|
| 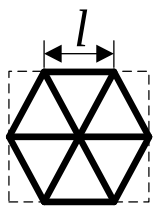 | $2\sqrt{3}\frac{b}{l}$ | $\frac{1}{3}\hat{\rho}$ | $\frac{1}{3}\hat{\rho}$ | $\frac{1}{8}\hat{\rho}$ | $\frac{1}{3}$ | $\frac{1}{3}$ |
| 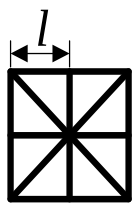 | $\left(2+\sqrt{2}\right)\frac{b}{l}$ | $\frac{2+\sqrt{2}}{5+3\sqrt{2}}\hat{\rho}$ | $\frac{2+\sqrt{2}}{5+3\sqrt{2}}\hat{\rho}$ | $\frac{1}{4\sqrt{2}+4}\hat{\rho}$ | $\frac{1}{2\sqrt{2}+1}$ | $\frac{1}{2\sqrt{2}+1}$ |
| 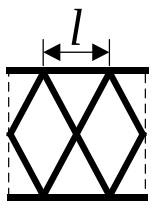 | $\frac{5}{\sqrt{3}}\frac{b}{l}$ | $\frac{1}{5}\hat{\rho}$ | $\frac{9}{25}\hat{\rho}$ | $\frac{3}{20}\hat{\rho}$ | $\frac{3}{5}$ | $\frac{1}{3}$ |

Furthermore, the FEM is employed to verify the theoretical prediction (more details can be found in Appendix A). Here the sizes of the components are chosen as $l/t$=10 and $b/t$=1. The stiffness coefficients $A_{ij}$ are normalized as

$$\hat{A}_{ij} = \frac{A_{ij} l}{(EA)_{\mathrm{f}}}. \tag{5.11}$$

Fig. **6** presents the normalized stiffness coefficients of the three networks obtained from theoretical model, FEM simulation and the results of Wang and Mcdowell (2004). It can be found that the results agree well with each other, indicating the accuracy of our theoretical model.

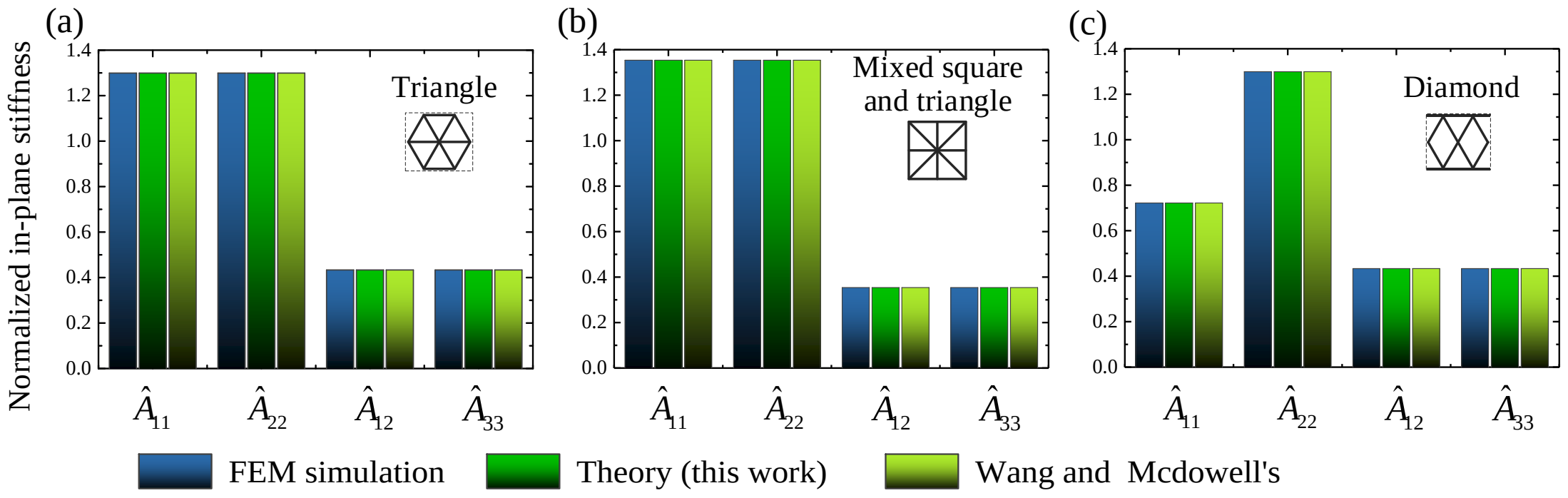

**Fig. 6** In-plane stiffness coefficients of (a) triangle, (b) mixed square and triangle and (c) diamond lattice networks. The results are obtained from FEM simulation, the theoretical model of this work and the model of Wang and Mcdowell (2004).

### *5.1.2 Out-of-plane stiffness*

Similarly, the out-of-plane stiffness of the three networks is also verified by FEM simulation. To make a comparison with the classical continuum plate model, an assumptive out-of-plane stiffness $\mathbf{D}^{\mathrm{c}}$ is derived from the in-plane stiffness and the thickness, as

$$D_{ij}^{\mathrm{c}} = \frac{t^2}{12} A_{ij}. \tag{5.12}$$

Besides, the normorlized stiffness is introduced here, as

$$\hat{D}_{ij} = \frac{l}{(EI)_{\mathrm{f}}} D_{ij} . \tag{5.13}$$

Note that the ratio $(GJ)_{\mathrm{f}}/(EI)_{\mathrm{f}}$ for the square sectional profile ($b/t$=1) is $0.141\times 6/(1+\nu)$ (Gere and Timoshenko, 1997), where the Poisson's ratio $\nu$ is taken as 0.3. Fig. 7 illustrates the normalized out-of-plane stiffness coefficients of the three networks, where the size and material parameters are the same as those in the cases in Fig. 6. The theoretical predictions are well verified by the FEM simulation. However, the assumptive out-of-plane stiffness $\mathbf{D}^{\mathrm{c}}$ of continuum plate model shows deviation from the present theoretical model and simulation. The stiffness $\mathbf{D}^{\mathrm{c}}$ of continuum plate model underestimates the coefficients $D_{11}$, $D_{22}$ and $D_{33}$ but overestimates the coupling coefficients $D_{12}$ ($D_{21}$), indicating that the out-of-plane stiffness cannot be derived from the in-plane stiffness and an effective thickness by the classical continuum plate model.

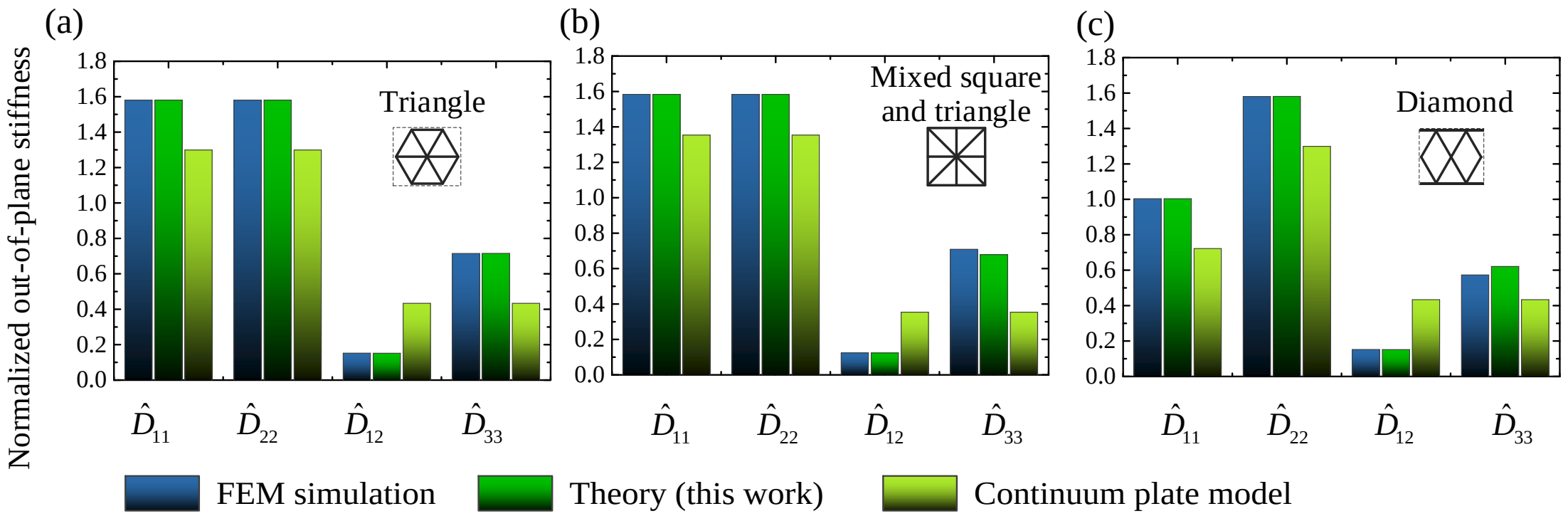

**Fig. 7** Out-of-plane stiffness coefficients of (a) triangle, (b) mixed square and triangle and (c) diamond lattice networks. The results are obtained from the theoretical model of this work, FEM simulation and continuum plate model.

## 5.2 Random networks

For randomly distributed networks, here we take an example as follows. All the components have the same length $l$. The position and orientation of the components are determined by the midpoint ($x_1$, $x_2$) and the angle $\theta$. In a given area $L_1 \times L_2$, $x_1$ and $x_2$ follow uniform distributions in the ranges of [0, $L_1$), [0, $L_2$) and $\theta$ complies with a uniform distribution function in the range of [-$\pi$, $\pi$), as

$$f(\theta)=\frac{1}{2\pi}. \tag{5.14}$$

Assuming the network is constructed by circular section filaments with a constant diameter $t$, the relative density of the network is rewritten as

$$\hat{\rho}=\frac{Nlt}{L_1L_2}, \tag{5.15}$$

and the total length of the components in the RVE is

$$l_{\text{total}}=Nl. \tag{5.16}$$

Substituting Eqs. (5.14)-(5.16) into Eqs. (4.27)-(4.38), the stiffness matrix can be derived as

$$\boldsymbol{A}=\frac{1}{8}\frac{\hat{\rho}}{t}(EA)_{\text{f}}\begin{bmatrix} 3 & 1 & 0 \\ & 3 & 0 \\ \text{sym} & & 1 \end{bmatrix}, \tag{5.17}$$

$$\boldsymbol{D}=\frac{1}{8}\frac{\hat{\rho}}{t}\begin{bmatrix} 3(EA)_{\text{f}}+(GJ)_{\text{f}} & (EA)_{\text{f}}-(GJ)_{\text{f}} & 0 \\ & 3(EA)_{\text{f}}+(GJ)_{\text{f}} & 0 \\ \text{sym} & & (EA)_{\text{f}}+(GJ)_{\text{f}} \end{bmatrix}. \tag{5.18}$$

Note that the in-plane stiffness is identical to Cox's model (Cox, 1952). However, for randomly distributed networks, there exist stiffness thresholds for the in-plane and out-of-plane stiffnesses, that is, only when the network density is larger than a critical value at which the load-transfer path is constructed, the stiffness becomes from zero to nonvanishing (Chen et al., 2015). Especially, for in-plane stiffness, only when the network density is larger than the "bending-stretching transitional threshold", can the components in the network carry load by axially stretching-dominated deformation. Therefore, according to the previous studies on the thresholds (Pan et al., 2016), the two matrixes should be modified as

$$\boldsymbol{A}=\frac{1}{8}\frac{\hat{\rho}-\left(\hat{\rho}_{\text{th}}^{\text{S1}}+\hat{\rho}_{\text{th}}^{\text{S2}}\right)/2}{t}(EA)_{\text{f}}\begin{bmatrix}3 & 1 & 0\\ & 3 & 0\\ \text{sym} & & 1\end{bmatrix}\quad\left(\hat{\rho}>\hat{\rho}_{\text{th}}^{\text{S2}}\right), \tag{5.19}$$

$$\boldsymbol{D}=\frac{1}{8}\frac{\hat{\rho}-\hat{\rho}_{\text{th}}^{\text{S1}}}{t}\begin{bmatrix}3(EA)_{\text{f}}+(GJ)_{\text{f}} & (EA)_{\text{f}}-(GJ)_{\text{f}} & 0\\ & 3(EA)_{\text{f}}+(GJ)_{\text{f}} & 0\\ \text{sym} & & (EA)_{\text{f}}+(GJ)_{\text{f}}\end{bmatrix}\left(\hat{\rho}>\hat{\rho}_{\text{th}}^{\text{S1}}\right) \tag{5.20}$$

where $\hat{\rho}_{\text{th}}^{\text{S1}}$ is the stiffness threshold of the rigidly connectied random network, and can be predicted by the electircal percolation thresold (Berhan et al., 2004; Pan et al., 2017), expressed as

$$\hat{\rho}_{\text{th}}^{\text{S1}}=\frac{5.8t}{l}, \tag{5.21}$$

and $\hat{\rho}_{\text{th}}^{\text{S2}}$ is the bending-streching transitional threshold, and it is predicted as

$$\hat{\rho}_{\text{th}}^{\text{S2}}=10.5\left(\frac{t}{l}\right)^{3/4}. \tag{5.22}$$

Here we only focus on networks with realtive large density, that is, the network with density larger than the bending-streching transitional threshold $\hat{\rho}_{\text{th}}^{\text{S2}}$.

The literature results and FEM simulation are also presented here. Fig. 8 shows the normalized in-plane stiffness coefficients of the random network with $L_1/l$ = 2.5 ($L_1 = L_2$) and $l/t$ = 400. To gain converged results in the FEM simulation, the result of each network density is calculated by the mean value of 80 networks with different random distributions. The component is simplified to be isotropic and linear elastic, same as the cases in Section 5.1. It can be found that the FEM simulation results agree well with the theoretical model in Eq.(5.19), and both the theoretical model and FEM simulation show a linear scaling law between the normalized stiffness and relative density. As a contrast, Cox's model (Cox, 1952) and Wu and Dzenis's model (Wu and Dzenis, 2005) cannot capture the threshold phenomenon although have a linear scaling relation.

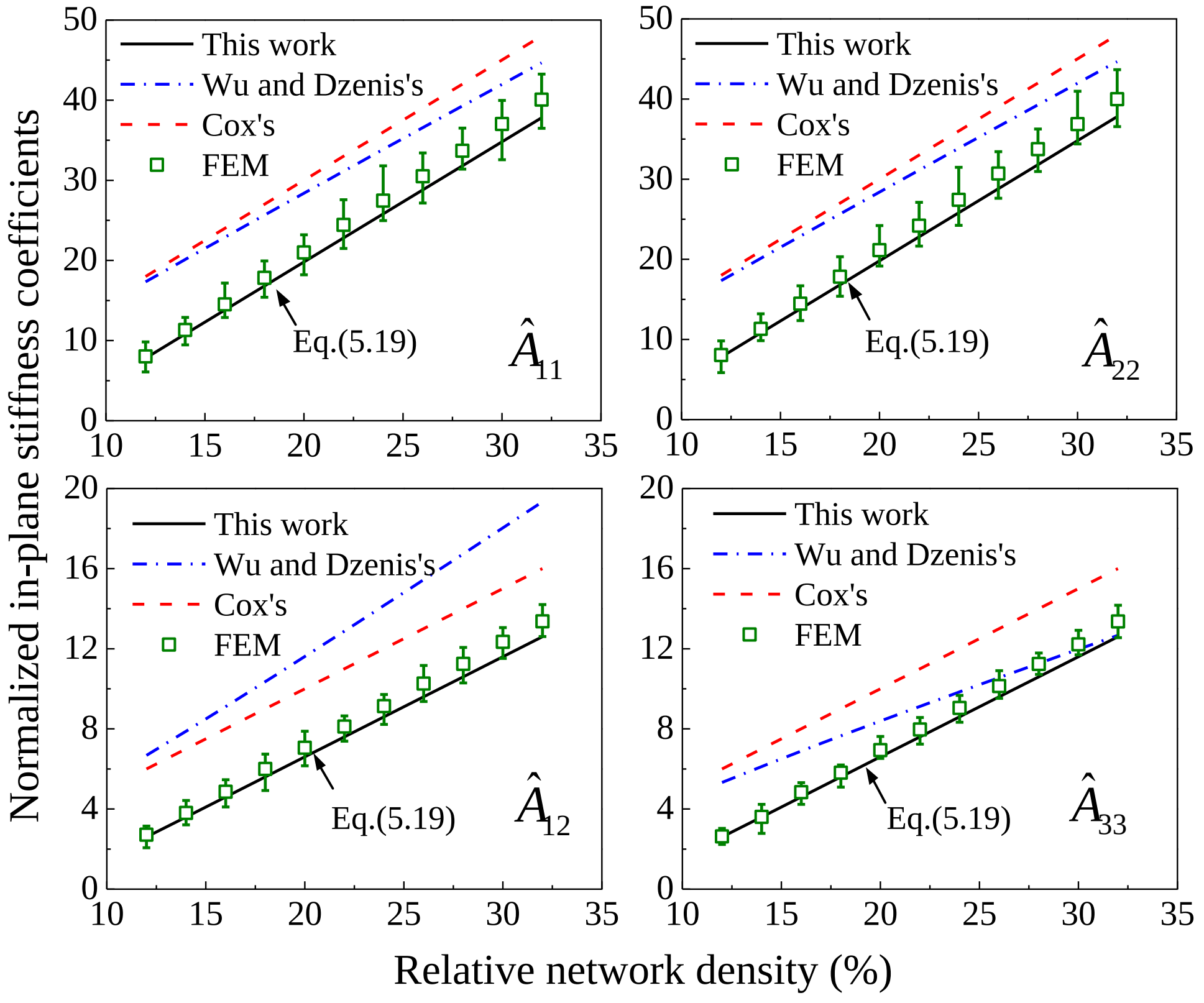

**Fig. 8** Curves of the normalized in-plane stiffness coefficients with respect to relative network density for random networks. The results are obtained from the theoretical model of this work, FEM simulation as well as the Cox's models (Cox, 1952) and Wu and Dzenis's models (Wu and Dzenis, 2005).

For the out-of-plane stiffness, the assumptive stiffness matrix derived from the in-plane stiffness and thickness according to classical continuum plate model is also employed to make a comparison. Here the ratio $(GJ)_{\mathrm{f}}/(EI)_{\mathrm{f}}$ for the circular sectional profile is $1/(1+\nu)$. As shown in Fig. 9, the scaling relations between the normalized out-of-plane stiffness coefficients and network density obtained from the theoretical model in Eq. (5.20) and FEM simulation can match well with each other. However, the assumptive stiffness derived from continuum plate model fails to predict the out-of-plane stiffness for the random networks.

Therefore, it can be concluded that the network model proposed in this work is well validated by FEM simulations and previous studies. The out-of-plane stiffness of all the presented networks cannot be predicted by the in-plane stiffness and thickness. The different deformation mechanism of the network is supposed to play the key role.

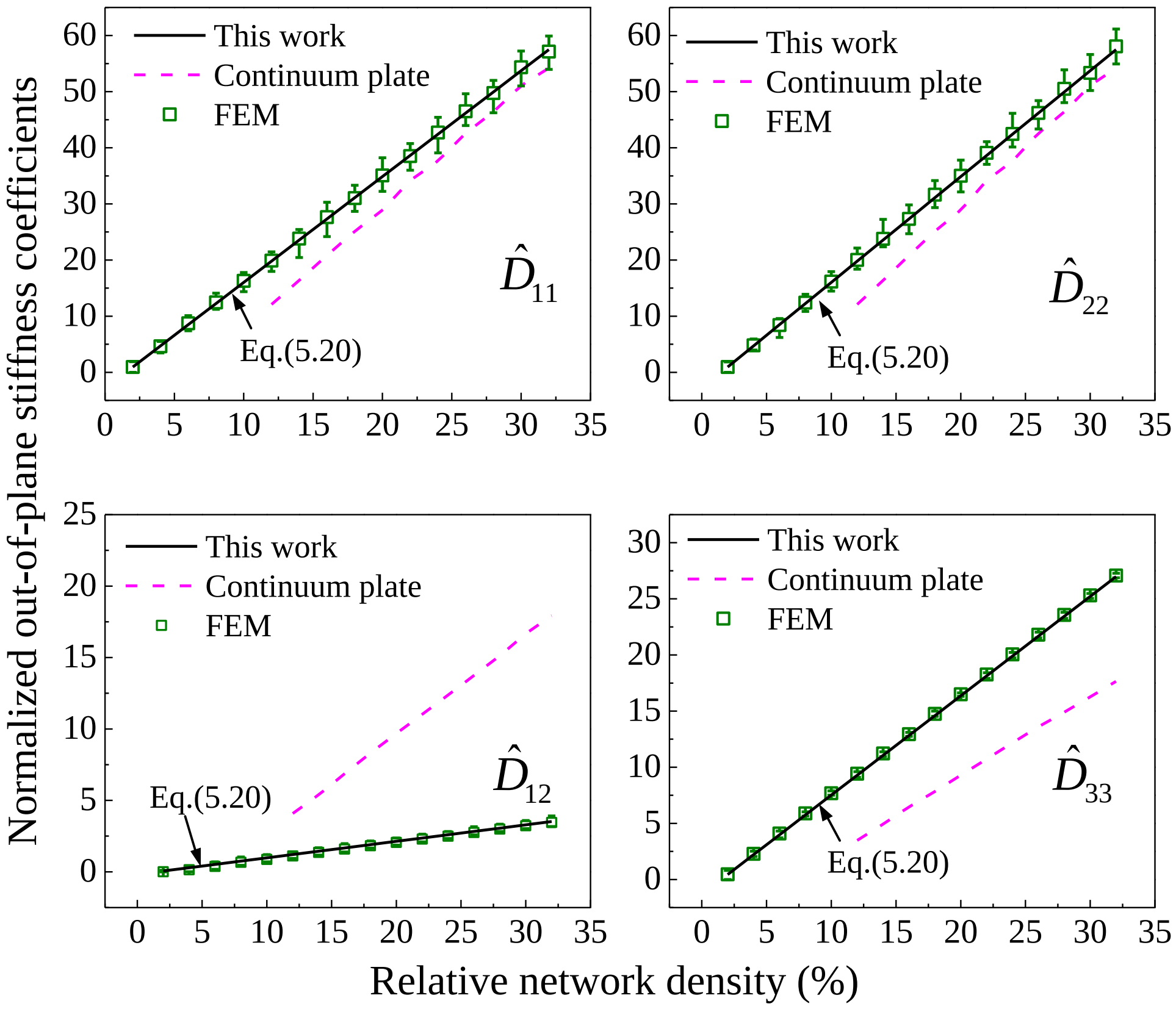

**Fig. 9** Curves of the normalized out-of-plane stiffness coefficients with respect to relative network density for randomly distributed networks. The results are obtained from the theoretical model of this work, FEM simulation as well as the classical thin plate model.

## 6 Discussion on the deformation mechanism

To make a deeper understanding of the deformation mechanism of the networked materials, the comparisons with continuum plate and 2D nanomaterial are carried out here respectively.

## 6.1 Thin networked materials vs. continuum plate

To make a comparison between the network and continuum plate, two micro units at the same position are taken out from the network and the continuum plate, where the one from the network is just a segment of the component, as shown in Fig. 10. The local strains and curvatures are projected to the surface of the units. For in-plane deformation, only the axial strain $\varepsilon_{xx}$ is nonzero and can contribute to the strain energy in the network, while all the strains (*i.e.* $\varepsilon_{xx}$, $\varepsilon_{yy}$ and $\varepsilon_{xy}$) in the continuum plate can induce strain energy. Similarly, in the out-of-plane deformation, the bending curvature $\kappa_{yy}$ is absent in the network due to the free lateral surface of the components. In the continuum plate, all the three curvatures can find a correspondence to an in-plane strain, and the one-to-one correspondence pairs are ($\varepsilon_{xx}$, $\kappa_{xx}$), ($\varepsilon_{yy}$, $\kappa_{yy}$) and ($\varepsilon_{xy}$, $\kappa_{xy}$). However, in the networked material, there is only one correspondence relation, that is ($\varepsilon_{xx}$, $\kappa_{xx}$), and $\kappa_{xy}$ cannot find any in-plane strain to build the correspondence. The correspondence relations in the continuum plate make the in-plane stiffness **A** and out-of-plane stiffness **D** have a simple linear relation, as Eq.(1.1), while no such relation can be established for networked materials due to the isolated torsion deformation $\kappa_{xy}$.

If the torsion deformation of the constitutive components is ignored, which means the torsional stiffness $(GJ)_{\mathrm{f}}$ is 0, and the stiffness threshold is neglected, it is interesting to find from Eqs. (4.15)-(4.38) that the out-of-plane stiffness of the network can degrade into

$$\mathbf{D}=\frac{I_{\mathrm{f}}}{A_{\mathrm{f}}}\mathbf{A}\,. \tag{6.1}$$

where $I_{\mathrm{f}}$ is the inertia moment and $A_{\mathrm{f}}$ is the sectional area of the constitutive components. Specifically, when the sectional profile is a rectangle, *i.e.* $I_{\mathrm{f}}/A_{\mathrm{f}}=t^2/12$, the networked materials have the same relation with the continuum plate, as Eq. (1.1). Therefore, taking the torsion deformation and the stiffness thresholds into consideration, the out-of-plane stiffness should be expressed as

$$\mathbf{D} = k_{\text{th}} \frac{I_{\text{f}}}{A_{\text{f}}} \mathbf{A} + \mathbf{D}^{\text{tor}} . \tag{6.2}$$

where $\mathbf{D}^{\text{tor}}$ is the contribution from the torsion deformation and can be obtained by only calculating the moments from the torsional torque $T_{\text{f}}$, or by setting $(EI)_{\text{f}} = 0$ in $\mathbf{D}$. For example, $\mathbf{D}^{\text{tor}}$ of random network can be obtained from Eq.(5.20), and expressed as

$$\boldsymbol{D}^{\text{tor}} = \frac{1}{8} \frac{\hat{\rho} - \hat{\rho}_{\text{th}}^{\text{S1}}}{t} \begin{bmatrix} (GJ)_{\text{f}} & -(GJ)_{\text{f}} & 0 \\ & (GJ)_{\text{f}} & 0 \\ \text{sym} & & (GJ)_{\text{f}} \end{bmatrix} \quad \left(\hat{\rho} > \hat{\rho}_{\text{th}}^{\text{S1}}\right). \tag{6.3}$$

$k_{\text{th}}$ is a modification coefficient to account for the threshold effect of the in-plane stiffness. For regular network, $k_{\text{th}} = 1$, and for random networks, it is expressed as

$$k_{\text{th}} = \frac{\hat{\rho} - \hat{\rho}_{\text{th}}^{\text{S1}}}{\hat{\rho} - \left(\hat{\rho}_{\text{th}}^{\text{S1}} + \hat{\rho}_{\text{th}}^{\text{S2}}\right)/2} . \tag{6.4}$$

The matrixes $\mathbf{A}$ and $\mathbf{D}^{\text{tor}}$ usually do not have a linear relation, so it is hard to find a constant effective thickness which is independent of loading modes to match the continuum plate model.

Therefore, it can be concluded that the anomalous relation between in-plane and out-of-plane stiffnesses of the networked materials mainly results from the isolated torsion deformation of the discrete structure, and for random networks, the asynchronous threshold effects between in-plane and out-of-plane stiffnesses also contributes to the anomalous relation.

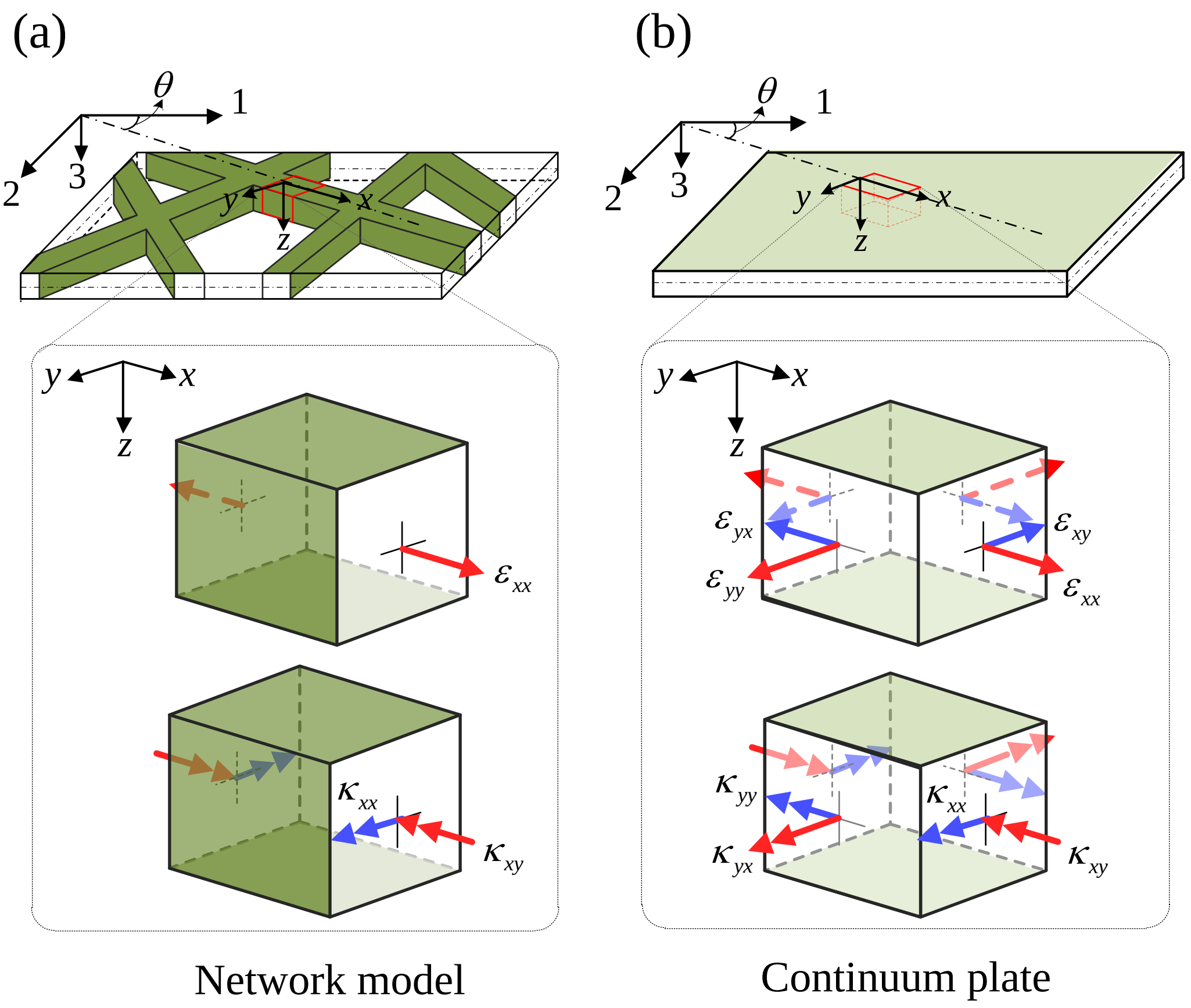

**Fig. 10** Scheme diagram for the deformation mechanism comparison between the (a) network model and (b) classical continuum plate model.

### 6.2 Thin networked materials vs. 2D materials in atomic scale

The mechanical behavior of 2D nanomaterials is determined by the inter-atomic interactions. Except for the weak long-range interactions like Van der Waals force, the interactions can be classified as bond, angle and dihedral angel typically (Chen et al., 2017; Rappé et al., 1992; Xu et al., 2013), as illustrated in Fig. 11a. The bond can be used to describe the distance variation between two atoms, which is corresponding to the stretching/compression deformation. The angle can be utilized to account for the in-plane or out-of-plane angle change for three adjacent atoms, which relates to bending, torsion or in-plane shearing deformations. Dihedral angle is usually used to describe the dihedral angle change between the two planes constructed by four atoms, which corresponds to bending and torsion deformations. Governed by these interactions, when under an external load, the atoms move from their initial position to a new equilibrium position. Therefore, the atom system can be modeled as beads and springs, as shown in Fig. 11b, in which the atom is rigid mass point, the bond is line spring and two types of angles are angular springs. The deformation of 2D materials actually is the deformation of these springs, and the strain energy is stored in these springs. Therefore, 2D materials have a totally different deformation mechanism from the continuum solids, and that is why they can resist bending deformation with only one layer atoms.

Besides, it can be found that there are some similarities between the atomic scale 2D materials and the networked materials, *i.e.* the discrete structures. From the view of the geometric feature, the intersections in the network are similar to the atoms, and the components are regarded as the bonds. The only difference is that, the “bonds” in atom system can only be stretched/compressed while in network can also be bent and twisted. However, some works have modeled the atom system as a beam framework, as shown in Fig. 11b, which is also known as molecular structural mechanics model (Kalamkarov et al., 2006; Li and Chou, 2003, 2004; Zaeri et al., 2010), and is very similar to the network model. It models the bonds between atoms as beams, which have stretching/compression, bending and torsion deformation modes, as shown in Fig. 11c. The mechanical properties of the beams in the molecular structural mechanics model need to be calibrated by the results of experiments or molecular simulations. Thus the geometric size or mechanical properties of the components can be assumed. Taking the graphene for example, the effective mechanical properties of the “bond” can be simplified as an isotropic beam with circular section, which has the diameter of cross section $d$ = 0.1466nm, Young’s mudolus $E$ = 5488 nN·nm$^{-2}$ and shear modulus $G$ = 871.1 nN·nm$^{-2}$ (Zaeri et al., 2010). Substituting these parameters into Eq.(4.21), the bending stiffness is obtained as 0.4196 nN·nm, which agrees well with the results obtained from molecular structural mechanics model 0.35-0.580 nN·nm (Berinskii et al., 2014; Safarian and Tahani, 2018; Shi et al., 2014). Therefore, combing with the molecular structural mechanics model, the network model developed in this work is also applicable in the atomic scale materials, which can help to give some insights into the mechanisms of the mechanical behavior of 2D nanomaterials.

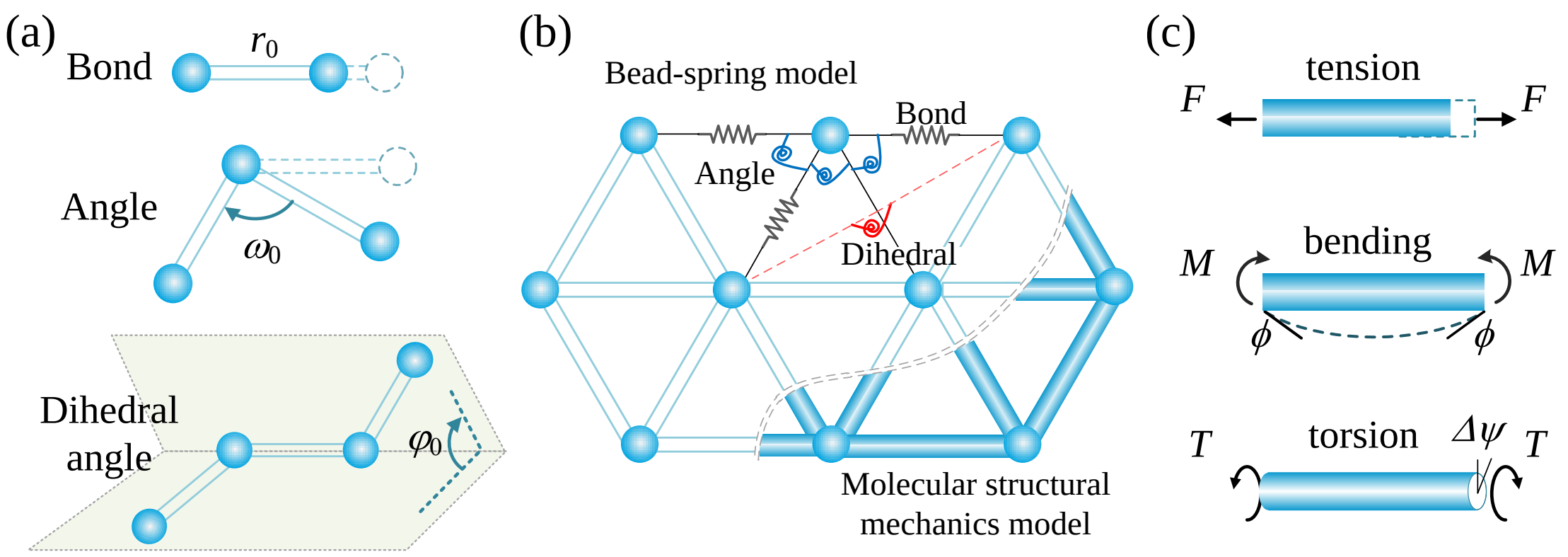

**Fig. 11** Scheme diagrams for (a) three typical inter-atomic interactions (*i.e.* bond, angle and dihedral angle), (b) bead-spring model and molecular structural mechanics model of atomic scale 2D materials and (c) equivalent beam for the "bond" between two atoms in the molecular structural mechanics model.

## 7 Conclusions

In this work, we have revealed the anomalous relation between in-plane and out-of-plane stiffnesses in thin networked materials with 2D discrete structural feature. Comparing to classical continuum plate model, the effective thickness of the networked materials obtained from their in-plane and out-of-plane stiffnesses presents a paradox. It cannot be properly defined by the relation in classical plate model, but varies with the loading modes, which is similar to the 2D materials in atomic scale. To reveal the mechanism underneath the paradox, we have established a micromechanical framework to investigate the deformation mechanism and predict the stiffness matrix of the networked materials.

By virtue of the theoretical modeling and FEM simulation, we have found the networks can carry external load by the axial stretching/compression, bending and torsion deformations of the components and captured the anomalous relation in some typical regular and random networks. Furthermore, It is revealed that, due to the discrete structure, there are many stress-free (moment-free) surfaces in the network, which break the correspondence relation between the in-plane strain and out-of-plane curvature in continuum solids, and make the torsion deformation isolated. The isolated torsional deformation then breaks the classical stiffness relation. Besides, the stiffness threshold effect in random network can further distort the relation between in-plane and out-of-plane stiffnesses. Based on these results, we have summarized a new formula to describe the stiffness relation. Finally, through the comparison between 2D materials in atomic scale, we found the network model also applies in atomic scale materials when combining with the molecular structural mechanics model. This work contributes to the in-depth understanding into the mechanical properties of 2D discrete materials/structures ranging from atomic scale to macro scale.

**Acknowledgments**

Supports from the National Natural Science Foundation of China (nos. 11622214, 11472027 and 11202012) and the Program for New Century Excellent Talents in University (no. NCET-13-0021) are gratefully acknowledged.

*Appendix A. Numerical implementation in FEM*

According to the periodic unit cell in Fig. 3 and Fig**.** 5, the FEM models are built, where the randomly distributed networks are generated by a pre-process code and with the size $L_1 = L_2$. To guarantee the convergence of the unit size of the randomly distributed networks, $L_1$ is taken to be larger than the length of the network components(Chen et al., 2015; Pan et al., 2017). The mechanical behaviors are simulated in the commercial FEM software ABAQUS/Standard (2005). The components of the network is modeled as rigidly connected beam and quadratic beam elements (B32 of Abaqus/Standard (2005)) are used in the simulation. For simplification, the material is isotropic and has a Poisson's ratio 0.3.

For in-plane loading, the periodic boundary conditions are applied on the unit cell by constraining nodes on opposite edges of the boundary (Chen and Ghosh, 2012; Segurado and Llorca, 2002; Wu and Koishi, 2009) as

$$u_\alpha^p - u_\alpha^{p'} = \varepsilon_{\alpha\beta}\left(x_\beta^p - x_\beta^{p'}\right), \quad (\alpha, \beta = 1, 2), \tag{A.1}$$

where $p$ and $p'$ are corresponding nodes of opposite edges of the unit cell, $\boldsymbol{\varepsilon}$ is strain tensor, $\boldsymbol{u}^p$ and $\boldsymbol{u}^{p'}$ are displacement of nodes $p$ and $p'$ respectively, and $\boldsymbol{x}^p$ and $\boldsymbol{x}^{p'}$ are the coordinates of nodes $p$ and $p'$ respectively. Besides, the rotation continuity of the boundary nodes on all the corresponding points is also taken into account as

$$\Delta\phi^p = \Delta\phi^{p'} . \tag{A.2}$$

For out-of-plane loading, the stiffness coefficients are obtained by applying uniaxial and biaxial bending to the network in the simulation. Due to the complex boundary condition of pure torsion deformation, the coefficients about the torsion are obtained by transformation of the stiffness matrix of a tilted unit cell. Taking the triangle lattice network for example, we can take two different periodic unit cells from the network, which are denoted as I and II, as shown in Fig. A1. Due to the symmetric geometry, the unit cell I (Fig. A1b) is orthotropic, *i.e.* ($D_{13}^{\mathrm{I}} = D_{23}^{\mathrm{I}} = 0$). According to tensor transformation (Reddy, 2004), the stiffness coefficients of the unit cell II satisfy

$$D_{11}^{\mathrm{II}} = D_{11}^{\mathrm{I}} \cos^4\alpha + 2\left(D_{12}^{\mathrm{I}} + 2D_{33}^{\mathrm{I}}\right)\sin^2\alpha\cos^2\alpha + D_{22}^{\mathrm{I}}\sin^4\alpha , \tag{A.3}$$

$$D_{12}^{\mathrm{II}} = \left(D_{11}^{\mathrm{I}} + D_{22}^{\mathrm{I}} - 4D_{33}^{\mathrm{I}}\right)\sin^2\alpha\cos^2\alpha + D_{12}^{\mathrm{I}}\left(\sin^4\alpha + \cos^4\alpha\right), \tag{A.4}$$

$$D_{22}^{\mathrm{II}} = D_{11}^{\mathrm{I}} \sin^4\alpha + 2\left(D_{12}^{\mathrm{I}} + 2D_{33}^{\mathrm{I}}\right)\sin^2\alpha\cos^2\alpha + D_{22}^{\mathrm{I}}\cos^4\alpha . \tag{A.5}$$

where $\alpha$ is the angle between the orientation of the two unit cells, as shown in Fig. A1a.

The coefficients related to bending, *i.e.* $D_{11}^{\mathrm{I}}$, $D_{22}^{\mathrm{I}}$, $D_{12}^{\mathrm{I}}$ and $D_{11}^{\mathrm{II}}$, $D_{22}^{\mathrm{II}}$, $D_{12}^{\mathrm{II}}$ can be obtained from the two periodic unit cells by uniaxial or biaxial bending simulation. Then, according to any of the relations in Eqs.(A.3)-(A.5), the coefficient $D_{33}$ can be obtained.

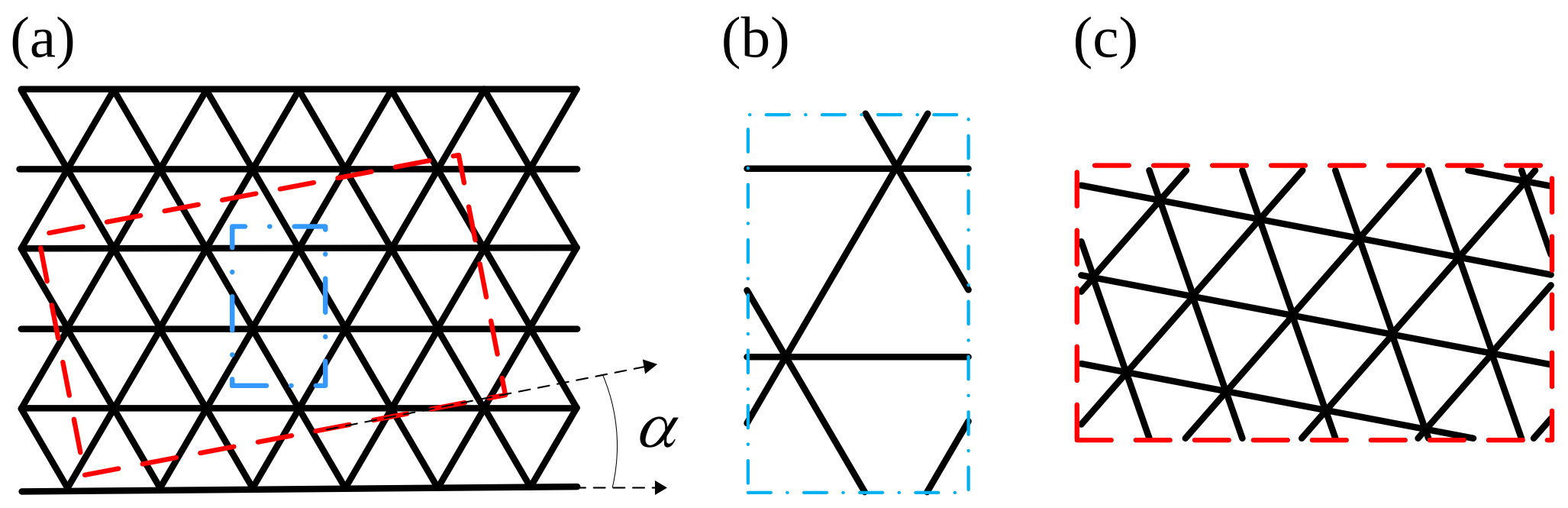

**Fig. A1** Scheme diagrams for (a) triangle lattice network, (b) periodic unit cell 1 and (c) periodic unit cell 2 that has an tilted angle $\alpha$.

The periodic boundary conditions are also applied on the unit cell by constraining nodes on opposite edges of the boundary. In this simulation, only uniaxial or biaxial bending is applied on the edges, and the boundary conditions are

$$u_i^p = u_i^{p'}, \quad (i = 1, 2, 3), \tag{A.6}$$

$$\Delta\phi_1^p - \Delta\phi_1^{p'} = \kappa_1 \left( x_1^p - x_1^{p'} \right), \tag{A.7}$$

$$\Delta\phi_2^p - \Delta\phi_2^{p'} = \kappa_2 \left( x_2^p - x_2^{p'} \right), \tag{A.8}$$

where $p$ and $p'$ are corresponding nodes of opposite edges of the unit cell, $\boldsymbol{u}^p$ and $\boldsymbol{u}^{p'}$ are displacement of nodes $p$ and $p'$ respectively, and $\boldsymbol{\phi}^p$ and $\boldsymbol{\phi}^{p'}$ are rotation angle of nodes $p$ and $p'$ respectively.

*Appendix B. Bending and torsion curvatures of the curve in a torsional plane.*

When a flat plane is under a small pure torsion $\kappa_3$, its shape can transform to a saddle surface, as

$$\mathbf{r}(x_1,x_2)=\left(x_1,x_2,-\frac{\kappa_3}{2}x_1x_2\right). \tag{B.1}$$

According to differential geometry (Kreyszig, 1991), the coefficients of the first fundamental form of the surface can be derived as

$$\underline{E}=1+\left(\frac{\kappa_3}{2}\right)^2 x_2^2, \tag{B.2}$$

$$\underline{F}=\left(\frac{\kappa_3}{2}\right)^2 x_1x_2, \tag{B.3}$$

$$\underline{G}=1+\left(\frac{\kappa_3}{2}\right)^2 x_1^2, \tag{B.4}$$

and the coefficients of the second fundamental form of the surface are

$$\underline{L}=0, \tag{B.5}$$

$$\underline{M}=\frac{-\frac{\kappa_3}{2}}{\sqrt{1+\left(\frac{\kappa_3}{2}\right)^2 x_1^2+\left(\frac{\kappa_3}{2}\right)^2 x_2^2}}, \tag{B.6}$$

$$\underline{N}=0. \tag{B.7}$$

The line on the surface can be determined by the starting point $\left(x_1^0,x_2^0\right)$ and the angle $\theta$, and after deformation, the spatial curve on the saddle surface can be written as

$$\begin{cases} u=x_1^0+s\cos\theta \\ v=x_2^0+s\sin\theta \\ w=-\frac{\kappa_3}{2}uv \end{cases}, \tag{B.8}$$

where the parameter $s$ notes the arc length of the curve.

The bending curvature is taken as the normal curvature of the curve, *i.e.*

$$\kappa(\theta)=\underline{L}\left(\frac{du(s)}{ds}\right)^2+2\underline{M}\frac{du(s)}{ds}\frac{dv(s)}{ds}+\underline{N}\left(\frac{dv(s)}{ds}\right)^2$$
$$=\frac{-\kappa_3\sin\theta\cos\theta}{\sqrt{1+\left(\frac{\kappa_3}{2}\right)^2x_1^2+\left(\frac{\kappa_3}{2}\right)^2x_2^2}}, \qquad \text{(B.9)}$$

and the torsional curvature is taken as the geodesic torsion. *i.e.*

$$\tau(\theta)=\frac{1}{\sqrt{\underline{E}\underline{G}-\underline{F}^2}}\begin{vmatrix}\left(\frac{dv(s)}{ds}\right)^2 & -\frac{du(s)}{ds}\frac{dv(s)}{ds} & \left(\frac{du(s)}{ds}\right)^2\\ \underline{E} & \underline{F} & \underline{G}\\ \underline{L} & \underline{M} & \underline{N}\end{vmatrix}. \qquad \text{(B.10)}$$
$$=\frac{-\frac{\kappa_3}{2}\left(\cos^2\theta-\sin^2\theta\right)-\left(-\frac{\kappa_3}{2}\right)^3x_1^2\sin^2\theta+\left(-\frac{\kappa_3}{2}\right)^3x_2^2\cos^2\theta}{1+\left(\frac{\kappa_3}{2}\right)^2x_1^2+\left(\frac{\kappa_3}{2}\right)^2x_2^2}$$

When the torsional curvature $\kappa_3$ is small, we can ignore the high-order items in Eqs. (B.9) and (B.10), thus the bending and torsion curvatures of the curve can be rewritten as

$$\kappa(\theta)=-\kappa_3\sin\theta\cos\theta, \qquad \text{(B.11)}$$

$$\tau(\theta)=-\frac{\kappa_3}{2}\cos2\theta. \qquad \text{(B.12)}$$

Note that in the analysis of out-of-plane deformation of the constitutive component, the reference directions of the bending and torsion curvatures are opposite to those in the above analysis, thus the signs should be changed.